# Ultrafast Polarization Switching via Laser-activated Ionic Migration in Ferroelectric $CuInP_2S_6$

Jin Zhang[1,*], Kun Yang[1], Jianxin Yu[1], Huixia Fu[2], Zijing Ding[3,*], Xinghua Shi[1], Sheng Meng[4-5]

[1]*Laboratory of Theoretical and Computational Nanoscience, National Center for Nanoscience and Technology, Chinese Academy of Sciences, Beijing 100190, P. R. China*
[2]*Center of Quantum Materials and Devices & Chongqing Key Laboratory for Strongly Coupled Physics, Chongqing University, Chongqing 401331, P. R. China*
[3]*Beijing Computational Science Research Center, Beijing 100193, P. R. China*
[4]*Beijing National Laboratory for Condensed Matter Physics, and Institute of Physics, Chinese Academy of Sciences, Beijing 100190, P. R. China*
[5]*Songshan Lake Materials Laboratory, Dongguan, Guangdong 523808, P. R. China*

*Corresponding author: Jin Zhang (jinzhang@nanoctr.cn)
Zijing Ding (zding@csrc.ac.cn)

As a layered ferroelectric material, $CuInP_2S_6$ has garnered significant attention for its robust ferroelectric state and potential applications in memory devices. In this work, we demonstrate that with short laser pulses ultrafast reversible polarization switching within hundreds of femtoseconds can be achieved in ferroelectric $CuInP_2S_6$. Specifically, photoexcitation triggers collective ionic migration and ferroelectricity reversal in $CuInP_2S_6$, revealing a novel pathway to access different ferroelectric phases through optical excitation. Our findings indicate that laser pulses substantially alter the transition barriers, promoting ionic transport facilitated by the photodoping effect. This laser-induced ionic migration proves critical for enabling polarization transitions, offering a novel pathway to explore and control exotic quantum phases. These insights open exciting possibilities for manipulating ferroelectric states and electronic properties on an unprecedented ultrafast timescale.

**Keywords**：Ferroelectricity, $CuInP_2S_6$, photoinduced phase transition, ionic migration, time-dependent density functional theory

Ferroelectric materials hold great potential as dielectrics in information storage and functional device applications [1–5]. These materials possess spontaneous electric polarizations below critical temperatures, which can be switched by electric fields. Layered ferroelectric materials facilitate the fabrication of ultrathin ferroic structures through exfoliation without dangling bonds. Additionally, they offer clean interfaces with other two-dimensional (2D) materials that can be modulated with external electric fields, promising for innovative electronic devices including tunnel junctions and ferroelectric field-effect transistors [6–11].

Recently, the study of ferroelectricity in 2D materials has opened up the possibilities for large-scale nano-electronic applications [12–15]. The family of van der Waals layered ferroelectric materials, including $CuInP_2S_6$ (CIPS) and $In_2Se_3$, are advantageous for controlling dielectrics for nonvolatile memory, ferroelectric field-effect transistors and advanced sensors [8,17-26]. In particular, CIPS is a representative 2D ferroelectric material wherein the Cu and In atoms displace in opposite directions within each layer [8-13]. It has long been recognized for its significant ionic conductivity at elevated temperatures, primarily ascribed to ionic hopping facilitated by the coupling between ionic vibrations and lattice deformation. In addition, Sokrates and coauthors revealed that CIPS hosts a quadruple-well potential with two distinct polar phases, attributing to the out-of-plane migration of Cu ions [12]. However, the relationship between ionic migration and carrier dynamics remains largely unexplored. The present work endeavors to bridge the gap by presenting a comprehensive investigation into the unique characteristics of CIPS, shedding light on the photoexcited lattice and carrier dynamics in ferroelectric materials.

In this Letter, we investigate photoexcited ultrafast polarization switching driven by ionic migration in the prototypical van der Waals ferroelectric CIPS, utilizing real-time time-dependent density functional theory (TDDFT) simulations (see more details in Supplementary Materials) [27]. These calculations allow us to unravel photoinduced structural changes at the atomic scale and femtosecond timescale. Optical excitation significantly modulates the ionic transition barriers in CIPS, triggering collective ionic dynamics. Furthermore, we identify intrinsic processes of polarization transitions, unveiling that various polarization states can be achieved through laser photoexcitation. This study not only provides profound insights into the ultrafast phase transitions in layered ferroelectric materials but also establishes a new framework for understanding photoinduced phenomena in a wide range of quantum materials.

Below the Curie temperature, the bulk CIPS displays a monoclinic structure with four formula units per unit cell, behaving as an insulator while exhibiting finite thermally assisted ionic conduction [12]. At higher temperatures (>42 °C), Cu ions become mobile and possibly occupy both intra- and interlayer lattice positions, which leads to a disordered state with no net polarization [10,13]. Previous studies have experimentally obtained the polarization of ferroelectric CIPS to be approximately 4 µC/cm$^2$ [22–24]. In contrast, when Cu ions in adjacent layers move in opposite directions, the interlayer stacking results in zero macroscopic polarization, corresponding to the antiferroelectric phase [13]. Despite its moderate polarization value, CIPS has recently been found to exhibit negative longitudinal piezoelectric coefficients. The inherent relationship between ferroelectricity and substantial ionic conductivity is intuitive, as long-range ion migration tends to disrupt dipole ordering. Further analysis indicates that applying hydrostatic pressure suppresses crystal distortion and forces Cu ions into interlayer sites, thereby increasing spontaneous polarization [15].

To estimate the thermal stability, we performed *ab initio* calculations on the total energies as a function of ionic displacements in CIPS, defined by the perpendicular distance between Cu ions and the adjacent sulfur layer at the bottom. As shown in Figure 1(a), the ferroelectric state exhibits an energy minimum at a displacement of 0.15 Å, confirming the ground-state configuration. When the displacement increases to 1.67 Å, the Cu ions move to the layer center, resulting in a state with no macroscopic polarization (*i.e.*, the paraelectric state). The energy barrier for this phase transition was determined using nudged elastic band optimization, where intermediate structures were linearly interpolated to locate the saddle point along the transition path. The barrier for the transition from the ferroelectric to the paraelectric state is 0.41 eV. For the configurations with negative displacements, Cu ions migrate into the interlayer positions or van der Waals gap, with a moderate barrier (0.13 eV) for the displacement of 0.78 Å.

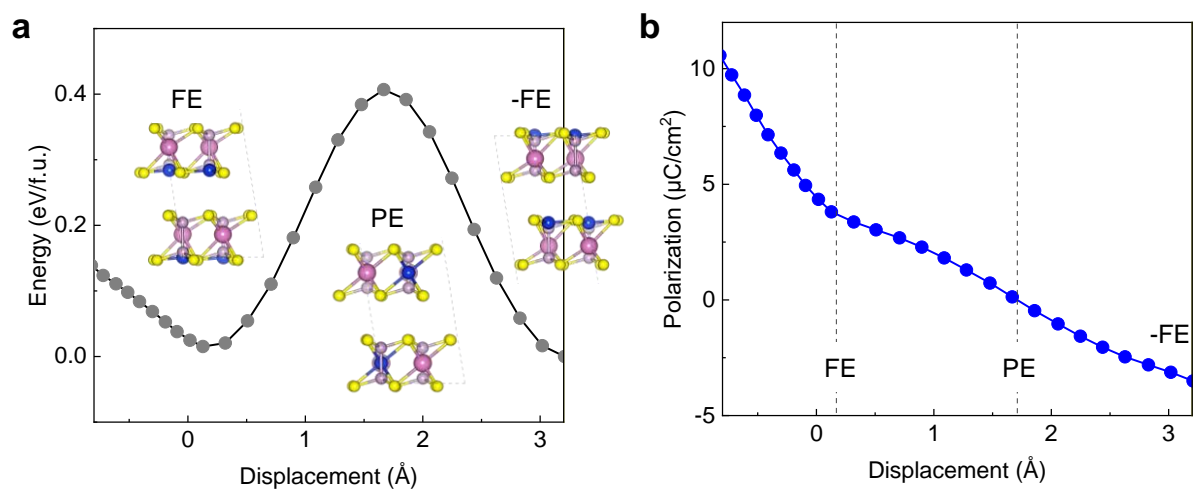

**Figure 1. Energy profile and total polarizations of CIPS at the ground state.** (a) Total energies with different ionic positions. Insets show the atomic structures of CIPS with the ferroelectric state (FE) and paraelectric state (PE). The state with inversed polarization is labeled with "-FE" in panel (a). The energy of the FE state is set to zero. (b) Polarizations with various Cu sites. The displacements of ferroelectric and paraelectric states are denoted in dashed lines. The energy barrier is abstained with the nudged elastic band optimization by linear interpolating along the transition path. The blue, red, purple and yellow spheres denote Cu, In, P and S atoms, respectively.

In our calculations, the ferroelectric CIPS phase exhibits a macroscopic polarization of 3.74 $\mu C/cm^2$, while the paraelectric state shows no net polarization, consistent with previous experimental and theoretical studies [22–24]. Notably, we find that the polarization doubles when Cu ions move into the van der Waals gap with a displacement of 0.78 Å. The displacement, together with the emergence of a state with large polarization (7.5 $\mu C/cm^2$), which resembles the high-polarization state although there is a difference in the lattice constants [12]. These findings reveal that polarizations are highly sensitive to ionic displacement. As exhibited in Figure 1(b), the net polarization depends directly on the ionic displacements in ferroelectric CIPS. This analysis establishes a clear relationship between the ionic displacement and macroscopic polarization.

Next, we conducted molecular dynamics (MD) simulations to examine the thermal effects and lattice propagation of CIPS under varying initial ionic temperatures. As demonstrated in Figure S1, the ionic displacement amplitude increases with rising temperatures. Snapshots of the final structures after 5 ps at different temperatures are also provided. At 300 K, the displacement amplitude of Cu ions grows to 1 Å. The Cu ions become delocalized into the interlayer space with an average displacement of 0.7 Å, indicating enhanced ionic delocalization without polarization switching even at elevated temperatures. This phenomenon is attributed to the out-of-plane phonon modes of the Cu ions with the phonon frequency at around 66 cm$^{-1}$.

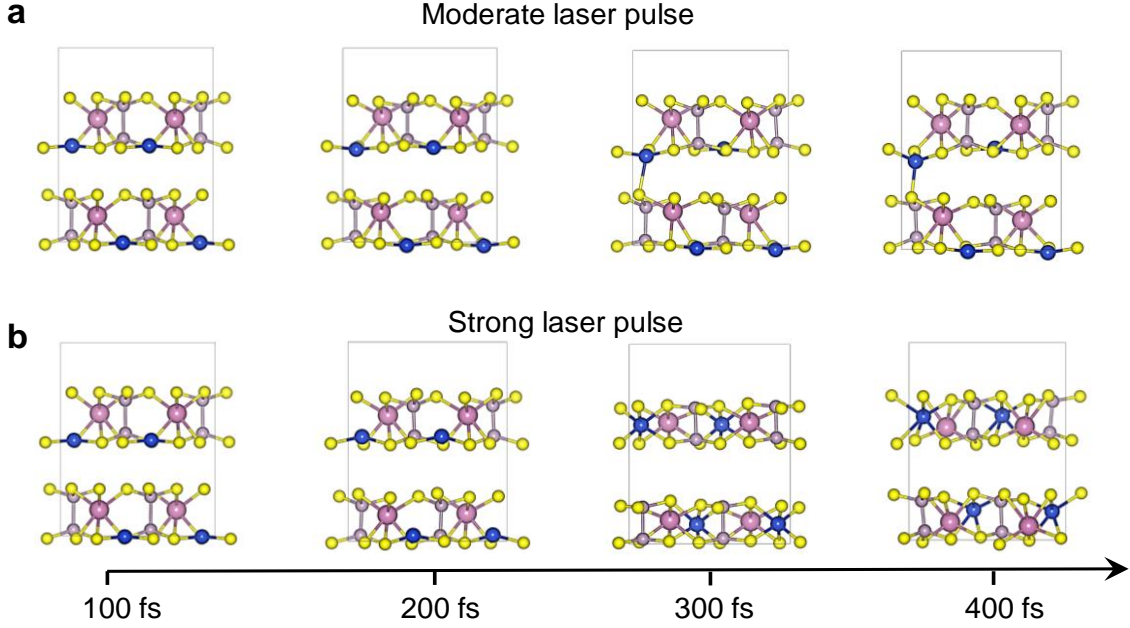

**Figure 2. Time evolution of atomic structures of bulk CIPS under photoexcitation.** (a) Snapshots for time-dependent atomic structures after photoexcitation at 0, 200, 400, and 600 fs for the photon energy of 3 eV with the laser intensity $E_0$=0.5 V/Å, respectively. (b) The same quantity as (a) with the laser intensity $E_0$=0.75 V/Å.

External electric fields are a conventional method for achieving polarization switching, but they face significant limitations, including the need for circuitry access and slow times on the nanosecond scale. Motivated by the above findings, we performed real-time TDDFT simulations to investigate laser-induced phase transitions in ferroelectric CIPS, capturing the intricate quantum behavior of the lattice and the evolution of excited states. As shown in Figure 2, the photoinduced structural changes are presented under different laser photon energies. To track laser-induced dynamics in bulk CIPS, a Gaussian-envelope function is used to describe the applied laser pulses.

$$E(t) = E_0 \cos(\omega t) \exp\left[-\frac{(t-t_0)^2}{2\sigma^2}\right], \qquad (1)$$

where $E_0$, $\omega$, $t_0$, and $\sigma$ are the maximum strength, the photon energy, the peak time of the electric field, and the width of the Gaussian pulse, respectively. The bandgap of CIPS is 1.45 eV on the level of the PBE functional, as illustrated in Figure S2. Laser pulses are employed with various photon energies to excite the system and the electronic and structural evolutions are monitored after photoexcitation. Upon laser irradiation, the electronic subsystem is directly excited, initiating selective excitations of various phonon modes through electron-phonon coupling.

We further investigate ultrafast photoinduced dynamics in CIPS under various laser frequencies and intensities. At a laser field strength $E_0$= 0.25 V/Å, a structural modulation is observed in ionic displacement with an amplitude of 1.0 Å after 500 fs for the photon energy of 3 eV. This displacement fluctuates around the value, but no polarization switching from ionic migration is detected. In contrast, at the moderate intensity $E_0$=0.50 V/Å, Cu ions at the bottom layer become significantly more mobile [Figure 2(a)]. After 400 fs, it is found that a subset of ions begins to migrate toward the interlayer space while the other Cu ions localize within the bottom sites with no obvious displacement.

In contrast, the ionic distribution undergoes complete reconfiguration due to photoexcitation for the strong laser field strength $E_0$=0.75 V/Å [Figure 2(b)]. All Cu ions migrate to the center of the intralayer space at t=300 fs and the ions are localized near the top layer by 400 fs, indicating that the polarization is fully switched by photoexcitation. We also observe lattice deformation, evidenced by a 28% reduction in layer height along the perpendicular direction at 300 fs, followed by a quick recovery. This finding sharply contrasts with the traditional understanding of polarization reversal, where changes in polarization induced by an electric field are predominantly attributed to ionic displacements.

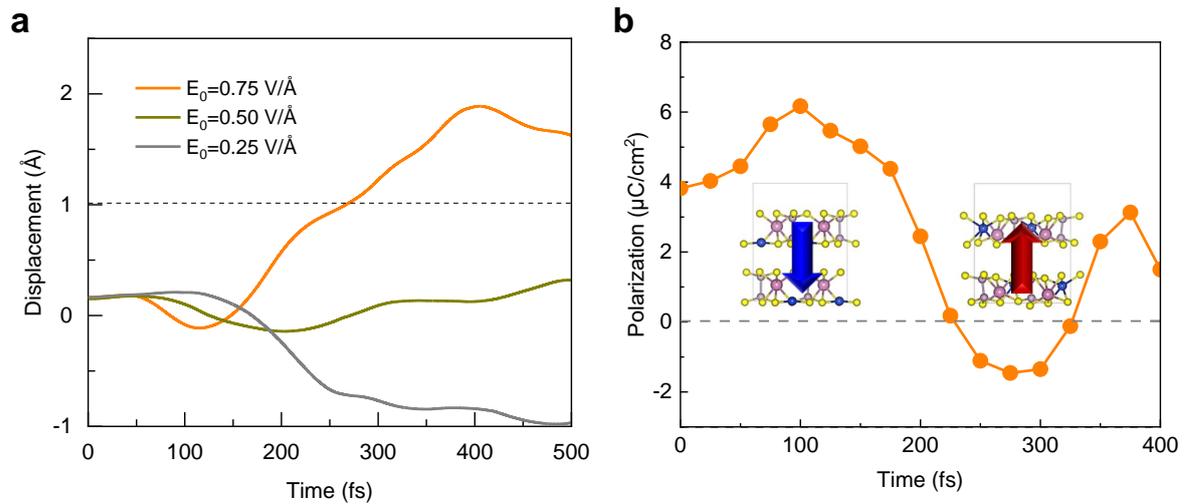

**Figure 3. Time evolution of atomic structure and polarization of CIPS under photoexcitation.** (a) Perpendicular displacements of Cu ions for the intensities ranging with the intensity from 0.25 to 0.75 V/Å. The applied photon energy is 3 eV. (b) Total polarizations during the evolution. The insets show the schematic phases and corresponding directions of total polarization. The dashed line denotes the states with no polarization with Cu ions at the middle of the intralayer spaces.

The time evolution of structures and polarizations under photoexcitation are shown in Figure 3(a). Specifically, the ionic displacement increases for the photon energy of 3 eV with the laser intensity $E_0$=0.75 V/Å, surpassing the ground-state configuration with all Cu ions localized at the bottom layer. The results unveil that the ferroelectric state is destroyed in an ultrafast timescale of 250 fs. The laser-induced dynamics in bulk CIPS differ significantly with the thermally-induced process. The ionic displacement is accompanied by a larger polarization, leading to the formation of a photoinduced ferroelectric state with Cu ions occupying the interlayer positions. The displacement reaches a peak of 0.32 Å at 450 fs, signaling the emergence of a new structural order. Notably, the oscillation is assigned to the phonon mode at the frequency of 66 cm$^{-1}$, corresponding to the out-of-plane mode of Cu ions. The results signify that lattice vibrations play a crucial role in ferroelectric polarization switching.

Figure 3(b) illustrates the dependence of total polarization on the laser-induced ionic displacement of Cu ions in CIPS. Our simulation exhibits the polariztion inctreased to 6.5 µC/cm$^2$, which is accoupanied with the Cu ions moving into the van der Waals gap for the first 100 fs after photoexcition. Furthermore, the Cu ions transport along the oppsise direction into the intralayer space and the detailed analysis shows that the total polarization decreases from 3.5 to 0 µC/cm$^2$ within 400 fs and reaches an opposite polarization of 2 µC/cm$^2$ at approximately 300 fs. Our findings demonstrate that laser pulses can significantly modulate ionic migration upon photoexcitation. The total polarization is fully switched within 400 fs, coinciding with the significant displacement of Cu ions.

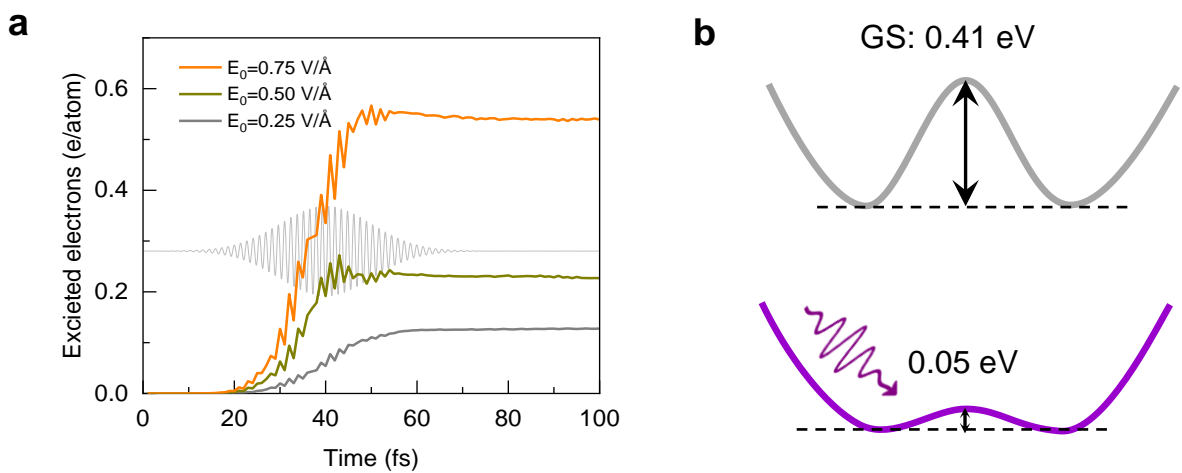

**Figure 4. Photoexcited carrier dynamics and ionic transition barriers in CIPS.** (a) The number of excited electrons upon photoexcitation from the valence to conduction bands under the laser pulses. The inset shows the shape of the applied electric field along the out-of-plane direction. The photon energy is 3 eV with the polarization along the perpendicular axis, and the shape of the applied laser pulse is shown in light grey. (b) Schematic profiles of the energy barriers at the ground state (GS) and photoexcited state, respectively. Photoexcitation can effectively lower the barrier of intralayer ionic transport.

Photoexcited carriers are generated by exciting electron-hole pairs under illumination. The concentration of photoinduced carriers can effectively modulate the ionic transition barriers. To understand the modulation of ionic transition barriers, the photodoping effect is systematically examined in CIPS, specifically the time evolution of excited carrier populations. Figure 4(a) shows the number of electrons excited from the valence to the conduction bands during laser excitation. The dependence of the number of excited electrons on the laser strength is calculated by projecting the time-evolved wavefunctions on the basis of the ground-state wavefunctions. Although ground-state wavefunctions depend on the time-propagated atomic coordinates, the self-consistent calculations are performed using the fixed atomic structures in the ground state as the configuration experiences negligible change in the early stage within several tens of femtoseconds. For the photon energy of 3 eV, we observe a doping carrier density of 0.13 e/atom with the intensity $E_0$= 0.25 V/Å. By contrast, the total density of excited carriers increases to 0.23 e/atom for 0.50 V/Å. Additionally, the excited carrier density rises to 0.54 e/atom for 0.75 V/Å. Our findings manifest that optical excitation contributes to an enhancement in charge density near the Cu ions and a noticeable depletion around the sulfur atoms. Therefore, the density of excited electrons increases with the strength of the laser field and stabilizes as the laser field diminishes.

To validate these findings, laser pulses with various photon energies are further compared [Figure S3]. Clearly, the laser pulse with a large energy of 7 eV introduces ferroelectric reversal in CIPS at the relatively small intensity of 0.25 V/Å. The displacement decreases for the high photon energy, surpassing the ground-state configuration, reflecting that the ferroelectric state is destroyed in an ultrafast timescale of 400 fs. By contrast, the laser-induced dynamics in bulk CIPS differ markedly for the photon energy of 5 eV. The Cu ions experience small fluctuations with the laser intensity $E_0$=0.25 V/Å, along the perpendicular direction due to the inherent lattice rigidity. The laser-induced displacement results in increased polarization and the formation of a photoinduced ferroelectric state, with Cu ions occupying interlayer positions. The results suggest that lattice vibrations play a crucial part in the polarization change in ferroelectric CIPS.

We also illustrate the corresponding lattice temperatures after photoexcitation for the large photon energy to understand the laser-induced changes, as shown in Figure S4. The effective temperature increases obviously due to the energy transfer from the Kohn-Sham electronic orbitals to the kinetic and static potential energy of the lattice subsystem. The concurrent ionic temperature shows that ultrafast polarization switching is not solely introduced by the lattice temperature. During this process, the lattice temperature spikes to above 500 K and then oscillates around this value. This indicates that the ultrafast transition of the ferroelectric state and ionic migration occurs when the lattice temperature exceeds its equilibrium thermal transition point (around 400 K). Our TDDFT-MD simulations naturally account for the damping of high-energy quasiparticles. These quasiparticles dissipate the energy to electrons at lower energy levels and the ionic subsystem. Therefore, excited electrons rapidly thermalize through electron scattering, forming a hot electron gas with an electronic temperature of several thousand Kelvin, significantly higher than that of the lattice subsystem.

It is important to note that the lattice temperature in a laser-induced nonequilibrium system differs from the equilibrium ionic temperature. The temperature comes from the kinetic energy of all ions defined as: $\tilde{T} = Mv_{ions}^2/3k_B$, where $v_{ions}$ is the ion velocity, M is the atomic mass and $k_B$ is the Boltzmann constant. The large oscillation in $\tilde{T}$ is interpreted as the kinetic energy and ionic potential energy exchange coherently. Due to the small unit cell in our simulations, the temperature lacks sufficient statistical sampling. Thus, the ionic temperature cannot be directly compared with the experimental temperature for the phase transition. It is important to note that our TDDFT-MD simulations naturally account for the damping of high-energy excitations, where energies are dissipated to lower-energy carriers, and the lattice subsystem. However, it is essential to recognize that full equilibrium between electrons and ions is not achievable due to the limitations of the current TDDFT-MD approach [28-29].

The structural and electronic dynamics obtained from first-principles calculations provide insights into the photoinduced behavior following optical excitations in CIPS on femtosecond timescales. Furthermore, we demonstrate that the excited carriers significantly modify the ionic transition barriers, as illustrated in Figure 4(b). The energy barrier for Cu ion diffusion between the top and bottom sulfur layers decreases to 0.05 eV for a photodoping level of 0.3 e/atom, one order of magnitude smaller than the value at the ground state (0.41 eV). This unveils that the application of intense lasers and ultrafast electron-electron scattering alters the potential energy surface, potentially driving cooperative atomic motions toward a new photoinduced ferroelectric state. The excited carriers play a pivotal role in the ferroelectricswitching. Therefore, the phase dynamics can be postulated to result from both the modification of the ionic transition barrier in CIPS and the laser-induced rise in the effective temperature.

Our calculations provide direct insights into the structural and electronic dynamics following optical excitation in CIPS at femtosecond scales. Under laser illumination, strong photoexcitation generates a high density of electron-hole pairs. The excited carriers thermalize among themselves, forming a hot electron gas with a temperature higher than that of the lattice subsystem. The lattice is subsequently heated by the hot electrons via effective electron-phonon interactions. The switching of the ferroelectric state occurs when the lattice temperature exceeds its equilibrium thermal transition temperature. Additionally, the polarization reversal is enhanced by modifications of the ionic transition barriers in CIPS, leading to cooperative atomic motions that drive the ferroelectric reversal. Photoexcited carriers induce coherent ionic motions and lattice oscillations, distinct from thermally excited dynamics and originate from energy transfer between the electronic and lattice subsystems. From a broader perspective, our results highlight the critical of ionic migration in ferroelectric switching, suggesting that photodoping may also significantly influence ionic migration and polarization switching in van der Waals ferroelectric materials.

In summary, *ab initio* TDDFT-MD simulations demonstrate the nature of laser-induced polarization changes and switching through ionic migration in ferroelectric CIPS, uncovering novel collective dynamics driven by the photodoping effect, distinctly different from thermally-induced polarization reversal. Our results prove that photodoping-modulated the ionic transition barriers, facilitate the ultrafast reversal of the ferroelectric state in the bulk CIPS. This work offers new insights into the photoinduced ferroelectric–paraelectric switching in van der Waals ferroelectric materials and suggests that the methodologies used here could be applied to a broad range of laser-modulated quantum materials, promising for applications including nonvolatile memory and ferroelectric field-effect transistors.

**Acknowledgments**

This work was supported by the starting funding from National Center for Nanoscience and Technology. This work was supported by the National Key R&D Program of China (2022YFA1203200), the Basic Science Center Project of the National Natural Science Foundation of China (22388101), the Strategic Priority Research Program of the Chinese Academy of Sciences (XDB36000000), the National Natural Science Foundation of China (12125202). The numerical calculations in this study were partially carried out on the ORISE Supercomputer. We thank fruitful discussions with Qing Yang.

**Author contributions**

J.Z. designed the research. All authors contributed to the analysis and discussion of the data and the writing of the manuscript.

**Conflict of Interest:** The authors declare no competing financial interest.

# Supplemental Materials for

# Ultrafast Polarization Switching via Laser-activated Ionic Migration in Ferroelectric $CuInP_2S_6$

Jin Zhang[1,*], Kun Yang[1], Jianxin Yu[1], Huixia Fu[2], Zijing Ding[3,*], Xinghua Shi[1], Sheng Meng[4-5]

[1]*Laboratory of Theoretical and Computational Nanoscience, National Center for Nanoscience and Technology, Chinese Academy of Sciences, Beijing 100190, P. R. China*

[2]*Center of Quantum Materials and Devices & Chongqing Key Laboratory for Strongly Coupled Physics, Chongqing University, Chongqing 401331, P. R. China*

[3]*Beijing Computational Science Research Center, Beijing 100193, P. R. China*

[4]*Beijing National Laboratory for Condensed Matter Physics, and Institute of Physics, Chinese Academy of Sciences, Beijing 100190, P. R. China*

[5]*Songshan Lake Materials Laboratory, Dongguan, Guangdong 523808, P. R. China*

*Corresponding author: Jin Zhang (jinzhang@nanoctr.cn)

Zijing Ding (zding@csrc.ac.cn)

**This file contains:**

Note 1. TDDFT methods and parameters

S1. Snapshots of Born-Oppenheimer MD at different Temperatures

S2. Atomic structure and electronic properties of CIPS

S3. Time evolution of structures and polarizations of CIPS under different photon energies

S4. Ionic temperature for CIPS after photoexcitation

**Note. TDDFT methods and parameters**

The calculations were performed utilizing the time-dependent *ab initio* package (TDAP), developed based on the time-dependent density functional theory [1-3] and implemented within SIESTA [4-6]. Below the Curie temperature, CIPS has a monoclinic structure consisting of four formula units per unit cell, with cell parameters a = 6.05 Å, b = 10.54 Å, c = 13.28 Å and β = 99.12°, from experimental measurements [7]. The Brillouin zone was sampled with a 4×4×2 k-point grids. The Troullier-Martin pseudopotentials and the adiabatic local density approximation were used for the exchange-correlation functional [8]. The dynamic simulations were carried out with an evolving time step of 50 as for both electrons and ions within a micro-canonical ensemble.

To explain the TDDFT methods more, we present more description on our methods based on time-dependent Kohn-Sham equations for coupled electron-ion motion. We can perform ab initio molecular dynamics (MD) for coupled electron-ion systems with the motion of ions following the Newtonian dynamics while electrons following the TDKS dynamics. The ionic velocities and positions are calculated with Verlet algorithm at each time step. When the initial conditions are chosen, the electronic subsystem may populate any state, ground or excited, and is coupled nonadiabatically with the motion of ions.

One drawback of the Ehrenfest dynamics is that the approach represents a mean-field theory of the mixed quantum-classical system, with forces on the ions averaged over many possible adiabatic electronic states induced by the ionic motion [3]. This approximation works well for situations where a single path dominates in the reaction dynamics, for the initial stages of excited states before significant surface crossings take place, or for cases where the state-averaged behavior is of interest when many electron levels are involved as in condensed phases. However, the approach has limitations when the excited states involve multiple paths, especially when state specific ionic trajectories are of interest. In such cases, Ehrenfest dynamics fails as it describes the nuclear path by a single average point even when the nuclear wave function has broken up into many different parts. In this work, the deficiency is not critical because the laser induced path dominates in the whole reaction dynamics. In addition, we have tested the dynamics by altering the initial conditions such as ionic temperatures and laser intensities to confirm the robustness of our conclusion.

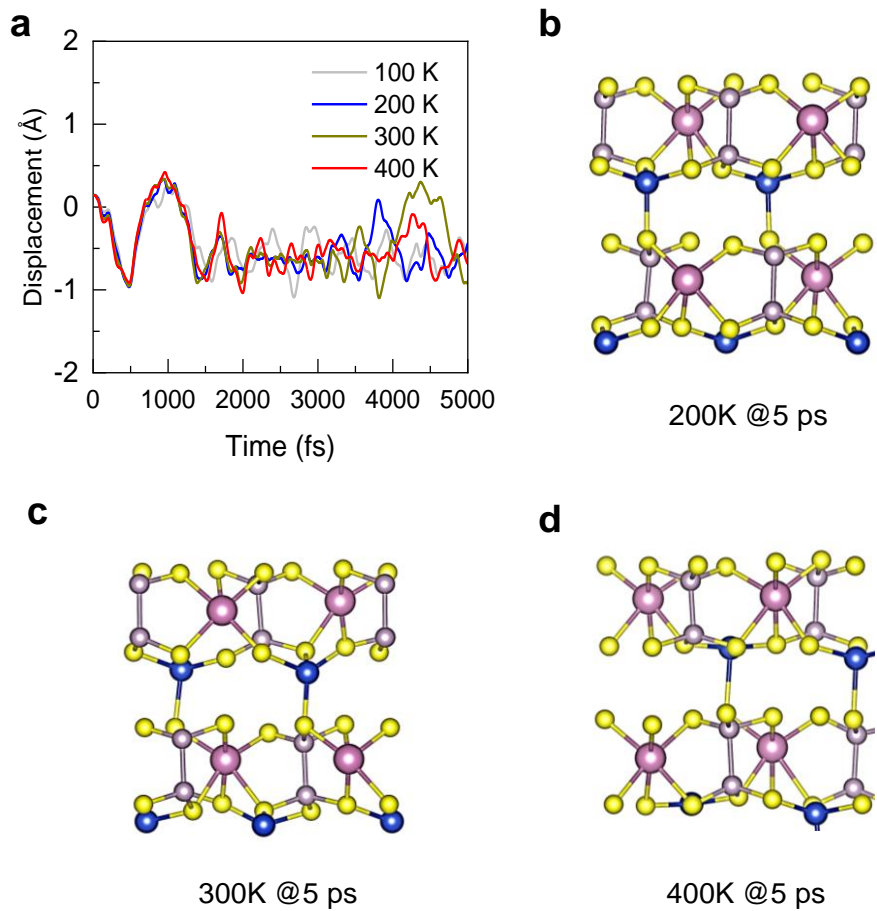

**Figure 1. Thermally-induced dynamics of CIPS.** (a) Displacements of Cu$^+$ ions in CIPS with initial temperatures from 100 to 400 K. The calculations are performed with Born-Oppenheimer molecular dynamics simulations. (b) Snapshot of the structures at 5 ps for 200 K from molecular dynamics simulations. (c) The same quantity as (b) for 300 K from molecular dynamics simulations.

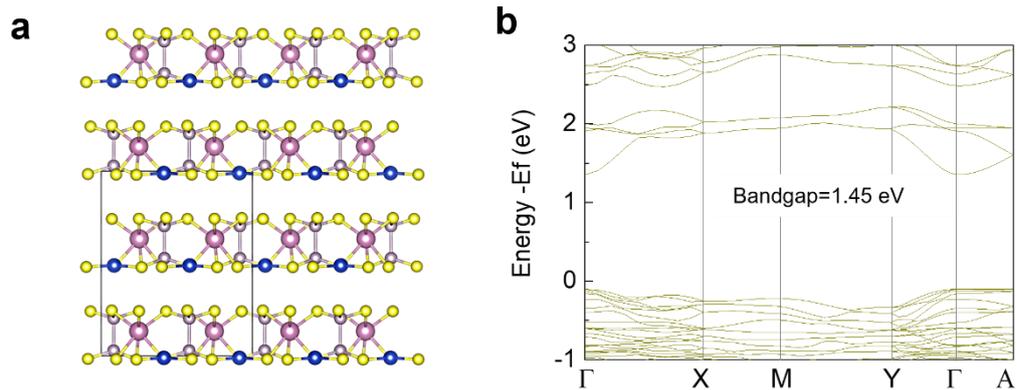

**Figure S2. Atomic structure and electronic properties of CIPS.** The unit cell shown with rectangle in (a). The bulk CIPS exhibits a direct bandgap of 1.45 eV on the level of the PBE functional (b).

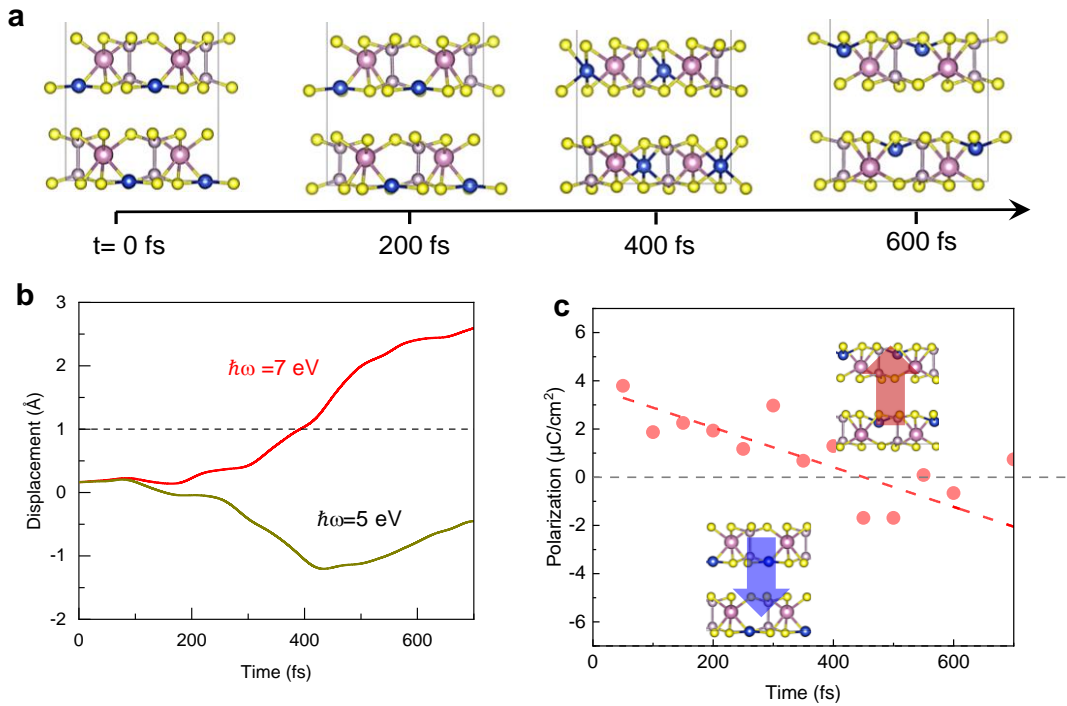

**Figure S3. Time evolution of structures and polarizations of CIPS under different photon energies.** (a) Snapshots for time-dependent atomic structures after photoexcitation at 0, 200, 400 and 600 fs for different photon energies, respectively. The photon energy is 7 Ev and the intensity is $E_0$ = 0.25 V/Å. (b) Corresponding perpendicular displacement of Cu ions for the photon energy of 7 eV and 5 eV, as shown in the inset. (c) Total polarizations during the evolution for the photon energy of 7 eV. The insets show the schematic phases and corresponding polarizations. The dashed line indicates the state where the Cu$^+$ ions occupy the center of interlayer sites of CIPS.

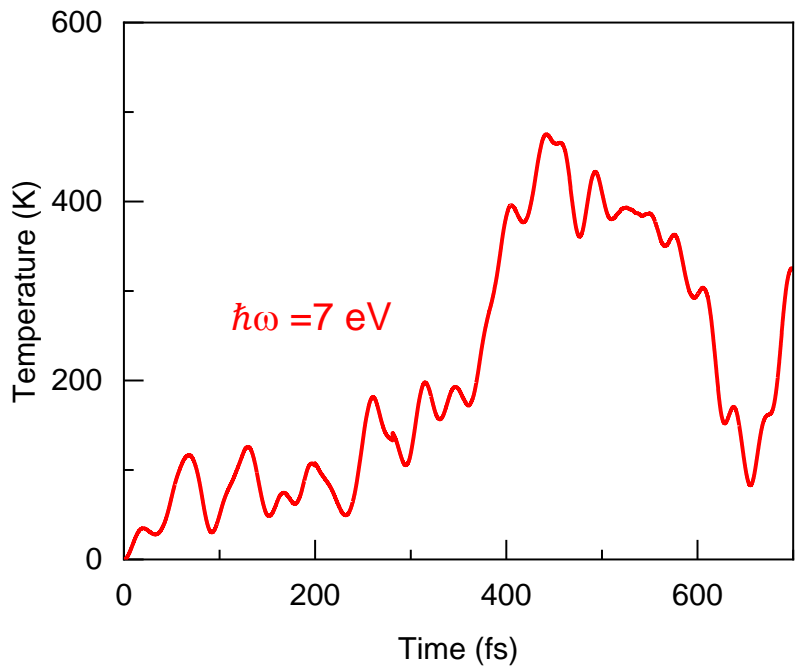

**Figure S4. Ionic temperature for CIPS after photoexcitation.** The photon energy is 7 eV and the intensity is E₀ = 0.25 V/Å. The temperatures are calculated from the kinetic energy of all ions defined as: $\tilde{T} = Mv_{ions}^2/3k_B$, where $v_{ions}$ is the ion velocity, M is the atomic mass and $k_B$ is the Boltzmann constant. The large oscillations in $\tilde{T}$ can be interpreted as the kinetic energy and ionic potential energy exchange coherently, while the small derivations at the later stage may be due to fluctuations.